\documentclass[usenatbib]{mn2e}
\usepackage{graphicx,tabulary,amsmath,amssymb,upgreek,wasysym,subfigure,overpic,array,hyperref}
\usepackage[T1]{fontenc}

\newcommand{\changes}[1]{{#1}}

\numberwithin{equation}{section}
\title[SPAMCART]{SPAMCART: a code for smoothed particle Monte Carlo radiative transfer}
\author[O. Lomax \& A. P. Whitworth]{O. Lomax\thanks{E-mail: oliver.lomax@astro.cf.ac.uk}, \& A. P. Whitworth\\
School of Physics and Astronomy, Cardiff University, Cardiff CF24 3AA, UK}

\begin{document}
\pagerange{\pageref{firstpage}--\pageref{lastpage}} \pubyear{2015}
\maketitle
\label{firstpage}

\begin{abstract}

We present a code for generating synthetic SEDs and intensity maps from Smoothed Particle Hydrodynamics simulation snapshots. The code is based on the \citet{Lucy99} Monte Carlo Radiative Transfer method, i.e. it follows discrete luminosity packets as they propagate through a density field, and then uses their trajectories to compute the radiative equilibrium temperature of the ambient dust. The sources can be extended and/or embedded, and discrete and/or diffuse. The density is not mapped onto a grid, and therefore the calculation is performed at exactly the same resolution as the hydrodynamics. We present two example calculations using this method. First, we demonstrate that the code strictly adheres to Kirchhoff's law of radiation. Second, we present synthetic intensity maps and spectra of an embedded protostellar multiple system. The algorithm uses data structures that are already constructed for other purposes in modern particle codes. It is therefore relatively simple to implement.

\end{abstract}

\begin{keywords}
radiative transfer -- hydrodynamics -- methods: numerical -- ISM: dust
\end{keywords}

\section{Introduction}

%
%
%
%

Monte Carlo Radiative Transfer (MCRT) is a stochastic method for simulating radiative transfer through a medium. Individual MCRT calculations are accurate but slow. Fortunately, they are trivial to parallelise and therefore well suited to modern multi-threaded CPUs. MCRT is often used to post-process simulation snapshots and can be adapted to solve a variety of radiative processes. For example, codes such as \textsc{hyperion} \citep{R11} and \textsc{radmc-3d} \citep{D12} calculate the dust and molecular line emissivities. These data can then be used to generate realistic synthetic observations. Other MCRT codes include \textsc{mocassin} \citep{EBSL03}, which models photoionisation fronts and emission line intensities from H\textsc{ii} regions, and \textsc{torus} \citep{H11}, which models time-dependent radiative transfer.

Smoothed particle hydrodynamics (SPH) \citep{L77,GM77} is a mesh-free method of solving the equations of fluid dynamics. A discrete set of particles are used to model the density distribution. These particles are smoothed over one another using a kernel function with smoothing length $h$. By letting $h$ vary with density, the particles can model a field spanning many orders of magnitude in density. This property makes SPH an attractive scheme for modelling astrophysical systems such as star forming clouds \citep[e.g.][]{B09a,LWHSW14,LWHSW14b}, galactic discs \citep[e.g.][]{DBP06,DBP11} and the cosmic web \citep[e.g.][]{SH03}. Open-source SPH codes include \textsc{gadget-2} \citep{S05} and \textsc{gandalf} \citep{HR13}.


Performing MCRT calculations on density fields from SPH simulations usually involves mapping the particles onto an \changes{octree} \citep[e.g.][]{SW05,R11,RHAB10}. This fundamentally changes the structure of the density field by adding noise and/or merging several particles into a single cell \changes{(we provide a brief analysis of this noise in Appendix \ref{apn:gridding})}. Another option is to use the particle positions to construct a Voronoi tessellation \citep[e.g.][]{HED16}. This is less noisy than an \changes{octree}, although the implementation is more complicated. Nevertheless, both methods alter the SPH density structure by replacing smoothed particles with uniform density cells.

In this paper we present the new code \textsc{spamcart} (short for Smoothed PArticle Monte CArlo Radiative Transfer). The code performs MCRT calculations using the properties of the particles' smoothing kernels. Therefore, post-processing calculations are performed at \emph{exactly} the same resolution as the SPH simulation. The algorithm is similar to other ray-tracing and Monte Carlo methods used in SPH, \citep[e.g.][]{ACP08,FR10b}. However, this is the first time such an algorithm has been used to perform full MCRT calculations. The method utilises the pre-existing neighbour-finding/gravity tree found in modern SPH codes, and is therefore relatively simple to implement.

In \S\ref{sec:MCRT} we describe the general Monte Carlo method we use for radiative transfer. In \S\ref{sec:spamcart} we explain how this method can be applied to smoothed particles. In \S\ref{sec:examples} we present some example calculations, including a physical benchmark. In \S\ref{sec:future} we list some future implementations and in \S\ref{sec:summary} we present the summary.


\section{Monte Carlo radiative transfer}
\label{sec:MCRT}

The method presented here is an adaptation of the \citet{Lucy99} algorithm, modified to operate on smoothed particles rather than a grid of uniform density cells. Most of the operations involve drawing a random variate $\mathcal{U}[a,b]$ from the Uniform distribution in the interval $(a,b)$.

\subsection{Luminosity packet trajectories}

A radiation source with luminosity $L$ emits $N_\gamma$ luminosity packets. Each packet is emitted from an origin $\boldsymbol{o}$ in a direction $\boldsymbol{n}$. If the radiator is an isotropic point source, $\boldsymbol{n}$ is given by polar angles
\begin{equation}
  \begin{split}
    \phi&=\mathcal{U}[-\uppi,\uppi]\,,\\
    \cos\theta_\textsc{iso}&=\mathcal{U}[-1,1]\,.
    \label{eqn:isotropic}
  \end{split}
\end{equation}
If there is an external isotropic radiation field, $\boldsymbol{o}$ is a random point on a closed convex surface and $\boldsymbol{n}$ is given by polar angles
\begin{equation}
  \begin{split}
  \phi&=\mathcal{U}[-\uppi,\uppi]\,,\\
    \cos\theta_\textsc{sur}&=\sqrt{\mathcal{U}[0,1]}\,,
  \end{split}
\end{equation}
where $\theta_\textsc{sur}$ is the angle between $\boldsymbol{n}$ and the inward surface normal. Each packet has energy $\varepsilon_0\equiv(L\,\Delta t/N_\gamma)$\footnote{If we assume radiative equilibrium, then the value of $\Delta t$ is arbitrary and does not affect the calculation.} and wavelength $\lambda$, drawn randomly from the spectral energy distribution of the source. For the surface of a star, approximated as a blackbody, the equation
\begin{equation}
  \frac{\int\limits_{0}^{\lambda} B_{\lambda'}(T)\,\mathrm{d}\lambda'}{\int\limits_{0}^{\infty} B_{\lambda'}(T)\,\mathrm{d}\lambda'}=\mathcal{U}[0,1]\,
\end{equation}
is solved for $\lambda$, where $T$ is the temperature of the source.

An individual packet travels an optical depth
\begin{equation}
  \tau=-\ln(\mathcal{U}[0,1])\,.
\end{equation}
This corresponds to a distance $l$, where
\begin{equation}
  \tau=\int\limits_0^l\rho(l')\,\chi_\lambda(l')\,\mathrm{d}l'\,.
  \label{eqn:optical_depth}
\end{equation}
Here, $\rho$ is the density of the medium and $\chi_\lambda$ is the mass extinction coefficient. Solving Eqn. \ref{eqn:optical_depth} for $l$ is one of the more computationally demanding aspects of MCRT and is covered in the next section.

Once a packet has travelled distance $l$, it is either absorbed or scattered at position
\begin{equation}
  \boldsymbol{o}_\textsc{new}=\boldsymbol{x}+l\,\boldsymbol{o}\,.
\end{equation}
A packet is absorbed if
\begin{equation}
  a_\lambda <\mathcal{U}[0,1]\,,
\end{equation}
where $a_\lambda$ is the albedo of the medium. Otherwise the packet is scattered. An absorbed packet is immediately re-emitted with a new wavelength drawn from the emissivity distribution of the medium. This is achieved by solving the equation
\begin{equation}
  \frac{\int\limits_{0}^{\lambda} \kappa_{\lambda'}\,B_{\lambda'}(T)\,\mathrm{d}\lambda'}{\int\limits_{0}^{\infty} \kappa_{\lambda'}\,B_{\lambda'}(T)\,\mathrm{d}\lambda'}=\mathcal{U}[0,1]\,
\end{equation}
for $\lambda$, where $\kappa_\lambda\equiv(1-a_\lambda)\,\chi_\lambda$ is the monochromatic mass absorption coefficient and $T$ is the temperature of the medium at $\boldsymbol{o}_\textsc{new}$. If the packet is scattered, $\lambda$ remains unchanged and $\boldsymbol{n}_\textsc{new}$ is scattered away from $\boldsymbol{n}$ by a random angle $\theta_\textsc{sct}$. Here, $\theta_\textsc{sct}$ is drawn from the \citet{HG41} phase function:
\begin{equation}
  \varPhi(\theta)=\frac{1}{4\,\uppi}\frac{1-g_\lambda^2}{[1+g_\lambda^2-2\,g_\lambda\,\cos(\theta)]^{3/2}}\,,
\end{equation}
where $g_\lambda$ is the mean scattering cosine. Hence, relative to the pre-scattering direction,
\begin{equation}
  \begin{split}
  \phi&=\mathcal{U}[-\uppi,\uppi]\,,\\
  \cos\theta_\textsc{sct}&=\frac{1}{2\,g}\left\{1+g^2-\left(\frac{1-g^2}{1+g\,\mathcal{U}[-1,1]}\right)^2\right\}\,.
  \end{split}
\end{equation}
The process of absorption then re-emission and/or scattering is repeated until the packet exits the system.

\subsection{Emissivity distribution}
\label{sec:sed}

For a medium in radiative equilibrium, the energy emission rate is equal to the energy absorption rate $\dot{A}$, i.e.
\begin{equation}
  \dot{A}=4\,\kappa_\textsc{p}\,\sigma_\textsc{sb}\,T^4\,.
  \label{eqn:rad_equil}
\end{equation}
Here, $\sigma_\textsc{sb}$ is the Stefan-Boltzmann constant, $T$ is the local temperature and $\kappa_\textsc{p}$ is the Planck mean absorption coefficient:
\begin{equation}
  \kappa_\textsc{p}=\frac{\int\limits_0^\infty \kappa_\lambda\,B_\lambda(T)\,\mathrm{d}\lambda}{\int\limits_0^\infty B_\lambda(T)\,\mathrm{d}\lambda}\,.
\end{equation}
The average value of $\dot{A}$ within a volume $V$ can be estimated by summing over the path lengths $l_j$ of luminosity packets which pass through $V$,
\begin{equation}
  \dot{A}\approx\frac{\varepsilon_0}{\Delta t}\frac{1}{V}\sum\limits_j \kappa_{\lambda_j} l_j\,\,.
  \label{eqn:rad_abs}
\end{equation}
We can estimate $T$, and therefore the mass emissivity distribution, by calculating $\dot{A}$ and solving Eqn. \ref{eqn:rad_equil} for $T$. However, Eqn. \ref{eqn:rad_abs} depends on the trajectories of the luminosity packets and the trajectories depend on the emissivity distribution. Therefore several iterations are required to reach an equilibrium solution. 

\section{SPAMCART algorithm}
\label{sec:spamcart}

The main algorithm of the \textsc{spamcart} code consists of two tasks: (i) to calculate the trajectories of luminosity packets through an ensemble of smoothed particles and (ii) to estimate $\dot{A}$ for each particle. Here, we detail how these tasks are performed.

\subsection{Density}

For an ensemble of smoothed particles, each with position $\boldsymbol{x}_i$, mass $m_i$ and smoothing length $h_i$, the density at any point in space is given by
\begin{equation}
  \rho(\boldsymbol{x})=\sum\limits_i \frac{m_i}{h_i^3}\,w(s_i)\,,\quad s_i=\frac{\lVert \boldsymbol{x}-\boldsymbol{x}_i\rVert}{h_i}\,,
\end{equation}
where $w(s)$ is the kernel function. Most kernel functions have compact support, i.e. they are only finite within $\xi$ smoothing lengths. We use the M4 cubic spline kernel \citep{ML85} which has $\xi=2$ (see Appendix \ref{apn:kernel}). The smoothing length of each particle is calculated so that
\begin{equation}
  \begin{split}
    h_i&=\eta\,\left(\frac{m_i}{\rho_i}\right)^\frac{1}{3}\,,\\
    \rho_i&=\sum\limits_j \frac{m_j}{h_i^3}\,w(s_{ij})\,,\quad s_{ij}=\frac{\lVert \boldsymbol{x}_j-\boldsymbol{x}_i\rVert}{h_i}\,.
  \end{split}
  \label{eqn:gather}
\end{equation}
These equations are solved by iteration and we normally adopt $\eta=1.2$ \citep[as suggested by][]{PM04}.

\subsection{Scatter Calculation}
\label{sec:scatter}

The value of some arbitrary quantity $Z(\boldsymbol{x})$ at an arbitrary position $\boldsymbol{x}$ can be estimated by \emph{scattering} the same quantity, from each particle to $\boldsymbol{x}$, via the kernel function:
\begin{equation}
  Z(\boldsymbol{x})=\sum\limits_i \frac{m_i}{\rho_i}\frac{Z_i}{h_i^3}\,w(s_i)\,.
\end{equation}
The gradient $\nabla Z(\boldsymbol{x})$ may also be estimated from the gradient of the kernel function:
\begin{equation}
  \begin{split}
    \nabla Z(\boldsymbol{x})&=\sum\limits_i \frac{m_i}{\rho_i}\frac{Z_i}{h_i^4}\,\nabla w(s_i)\,,\\
    \nabla w(s_i)=&\frac{\boldsymbol{x}-\boldsymbol{x}_i}{\lVert\boldsymbol{x}-\boldsymbol{x}_i\rVert}\,\frac{\mathrm{d}}{\mathrm{d}s} w(s_i)\,.
  \end{split}
  \label{eqn:scatter}
\end{equation}
The gradient of the M4 kernel function is given in Appendix \ref{apn:kernel}\,.

\subsection{Optical depth}
\label{sec:kernel}

\begin{figure}
  \includegraphics[width=\columnwidth]{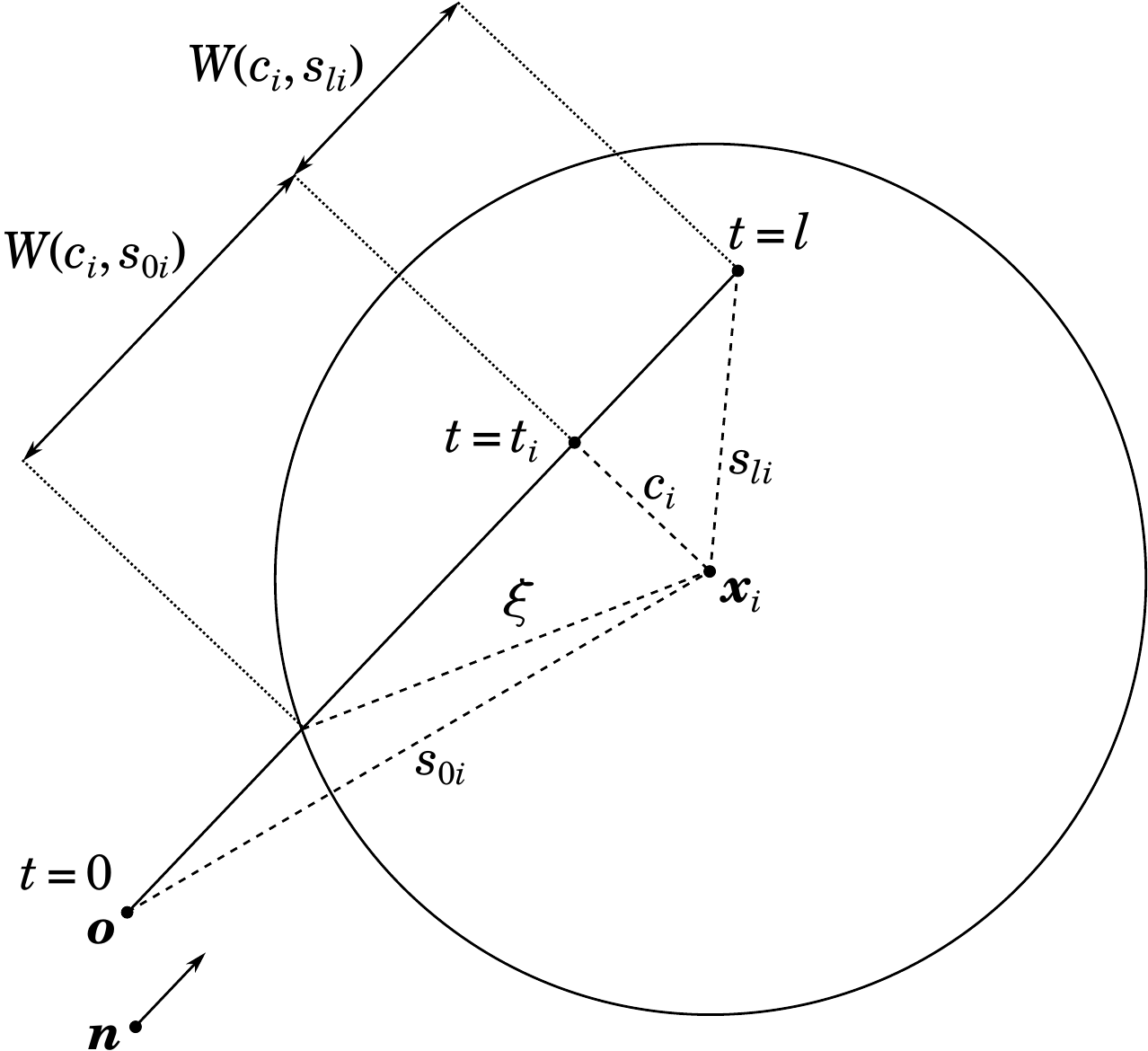}
  \caption{A diagram of line with origin $\boldsymbol{o}$, direction $\boldsymbol{n}$ and length $l$ intersecting a particle at position $\boldsymbol{x_i}$ with compact support $s\leq\xi$. The other terms are defined in \S \ref{sec:kernel}\,.}
  \label{fig:particle}
\end{figure}

The set of all points along a ray is defined by $\boldsymbol{x'}=\boldsymbol{o}+t\,\boldsymbol{n}$, where $0\leq t\leq l$. Here, $\boldsymbol{o}$ is the origin, $\boldsymbol{n}$ is the direction unit vector and $l$ is the length of the ray. The optical depth at wavelength $\lambda$ along the ray is given by
\begin{equation}
  \tau'(\boldsymbol{o},\boldsymbol{n},l)=\sum\limits_i^\textsc{ray}\chi_{\lambda i}\,\varsigma_i\,,
  \label{eqn:col_density}
\end{equation}
where
\begin{equation}
  \begin{split}
    \varsigma_i&=\frac{m_i}{h_i^2}
    \begin{cases}
      W(c_i,s_{0i})+W(c_i,s_{li}), & 0\leq t_i\leq l;\\
      |W(c_i,s_{0i})-W(c_i,s_{li})|, & \text{otherwise}\,,
    \end{cases}\\
    t_i&=(\boldsymbol{x}_i-\boldsymbol{o})\cdot\boldsymbol{n}\,,\\
    c_i&=\min\left(\frac{\lVert(\boldsymbol{o}+t_i\,\boldsymbol{n})-\boldsymbol{x}_i\rVert\,}{h_i},\xi\right)\,,\\
    s_{0i}&=\min\left(\frac{\lVert\boldsymbol{o}-\boldsymbol{x}_i \rVert}{h_i},\xi\right)\,,\\
    s_{li}&=\min\left(\frac{\lVert\boldsymbol{o}+l\boldsymbol{n}-\boldsymbol{x}_i \rVert}{h_i},\xi\right)\,,\\
    W(c,s)&=\int\limits_c^s\frac{w(s')\,s'}{\sqrt{s'^2-c^2}}\,\mathrm{d}s'\,,
  \end{split}
\end{equation}
and $\chi_{\lambda i}$ is the mass extinction coefficient of particle $i$. A diagram of this system for a single particle is given in Fig. \ref{fig:particle}\,. The analytical form of $W(c,s)$ for the M4 kernel in given in Appendix \ref{apn:kernel}.

\begin{figure}
  \includegraphics[width=\columnwidth]{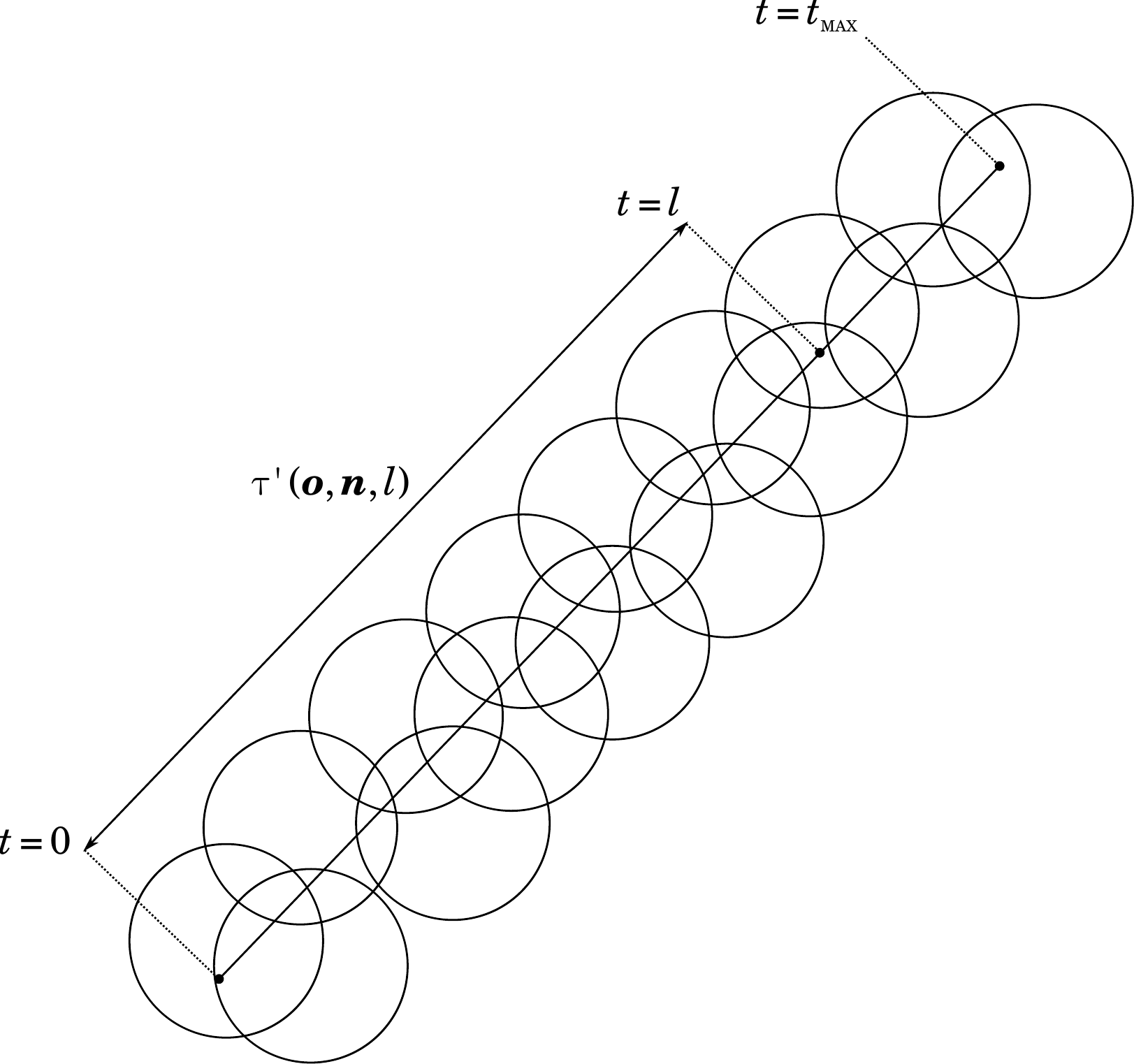}
  \caption{A diagram of \changes{all} particles which are intersected by a ray of length $t_\textsc{max}$. The optical depth along the ray may be calculated using Eqn. \ref{eqn:col_density}\,. Note that $t_\textsc{max}$ is greater than $l$. This is because $l$ is calculated by iteration and the first iterate must be an overestimate. \changes{Also, because of the construction of Eqn. \ref{eqn:col_density}, particle column density beyond $l$ does not contribute to the total optical depth.}}
  \label{fig:ray}
\end{figure}

In order to calculate Eqn. \ref{eqn:col_density}, we must first identify all the particles which are intersected by the ray (see Fig. \ref{fig:ray}). \changes{The sum of all particle column density contributions is then used to calculate the total optical depth. Particle-ray intersections can be identified} efficiently by walking a tree-structure (see Fig. \ref{fig:tree}) and opening cells which pass the slab test \citep[e.g.][see Appendix \ref{apn:tree}]{WBMS05}. These trees are a standard element of SPH codes, used to optimise neighbour-finding and gravity calculations.

\subsection{Propagating luminosity packets}
\label{sec:propogation}

\begin{figure}
  \includegraphics[width=\columnwidth]{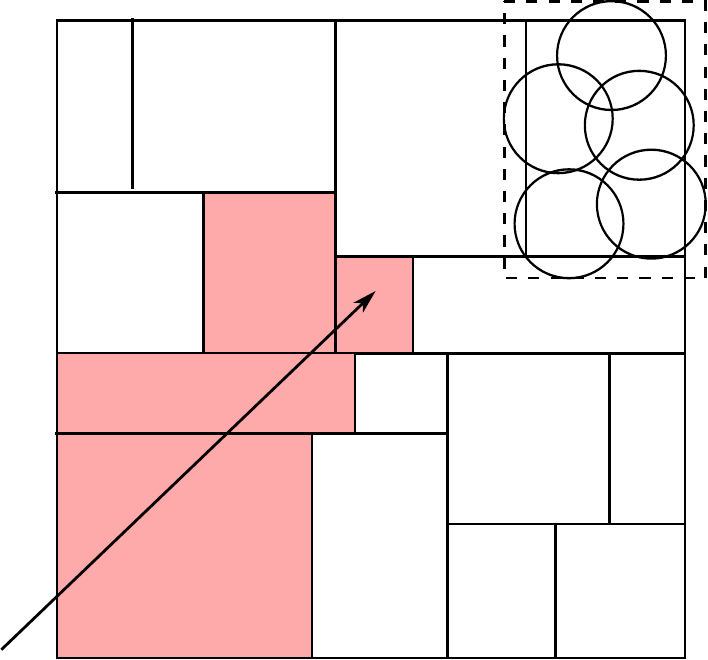}
  \caption{A schematic of the cells of a particle tree. Each cell has an axis-aligned bounding box (AABB) which encompasses the smoothing volume of all the particles in the cell (see Appendix \ref{apn:tree} for more details). If the ray intersects the AABB of a cell with sub-cells, the sub-cells are recursively checked. If the cell has no sub-cells (shaded pink), all of its particles are added to the ray.}
  \label{fig:tree}
\end{figure}

We propagate a luminosity packet, starting at $\boldsymbol{o}$, travelling in direction $\boldsymbol{n}$. Initially, the ray has length $l_0$. Here $l_0$ is a modest overestimate of the intended value of $l$ (see Eqn \ref{eqn:optical_depth}). A good choice of $l_0$ is provided by a second order Taylor expansion at $l=0$, i.e.
\begin{equation}
  \begin{split}
    l_0&=\frac{2\,\zeta\,\tau}{l^{-1}_\gamma(\boldsymbol{o})+\sqrt{\varDelta}}\,,\\
    \varDelta&=\max([l^{-1}_\gamma(\boldsymbol{o})]^2+2\,\tau\,\nabla l^{-1}_\gamma(\boldsymbol{o})\cdot\boldsymbol{n},0)\,.
  \end{split}
  \label{eqn:length_zero}
\end{equation}
Here, the inverse mean-free-path $l^{-1}_\gamma=\chi_\lambda\,\rho$ is estimated via a scatter calculation and $\zeta$ is an overestimation factor which we set to $1.2$\,.

We use a \changes{modified} Newton-Raphson root-finding method to iteratively solve $\tau'-\tau=0$ for $l$ \changes{(see Appendix \ref{apn:roots})}. We terminate the iterations when $|\tau'-\tau|/\tau\leq0.01$\,, \changes{usually within four or five iterations}. For some values of $l_0$, $\tau'$ is less than $\tau$ and no solution exists. Here, either the luminosity packet has left the ensemble of SPH particles or $l_0$ is too short. In the first case, the calculation for that luminosity packet is complete. In the second case, a new ray is constructed with the same direction and new origin $\boldsymbol{o}_\textsc{new}=\boldsymbol{o}+l_0\,\boldsymbol{n}$. The calculation is then repeated with \mbox{$\tau_\textsc{new}=\tau-\tau'$}.

\subsection{Estimating the energy absorption rate}

The energy absorption rate for an individual particle is estimated by summing the column densities along all packet trajectories:
\begin{equation}
  \dot{A}=\frac{\varepsilon_0}{\Delta t}\frac{1}{m}\sum\limits_j\kappa_{\lambda j}\,\varsigma_j\,.
  \label{eqn:dust_abs}
\end{equation}
Here $\varsigma_j$ is the column density along the path of luminosity packet $j$ through the particle. Estimating $\dot{A}$ requires almost no computational expense as the values of $\varsigma_j$ have already been computed whilst propagating the luminosity packets.

The absorption rate from the previous iteration is used when randomly generating a new emission wavelength. The value of $\dot{A}$ at an arbitrary position can be found via a scatter calculation. The radiative equilibrium calculation is complete when the change in $\dot{A}_i$ (or temperature $T_i$) between iterations is less than a desired tolerance.

\subsection{Generating intensity maps}

\begin{figure}
  \includegraphics[width=\columnwidth]{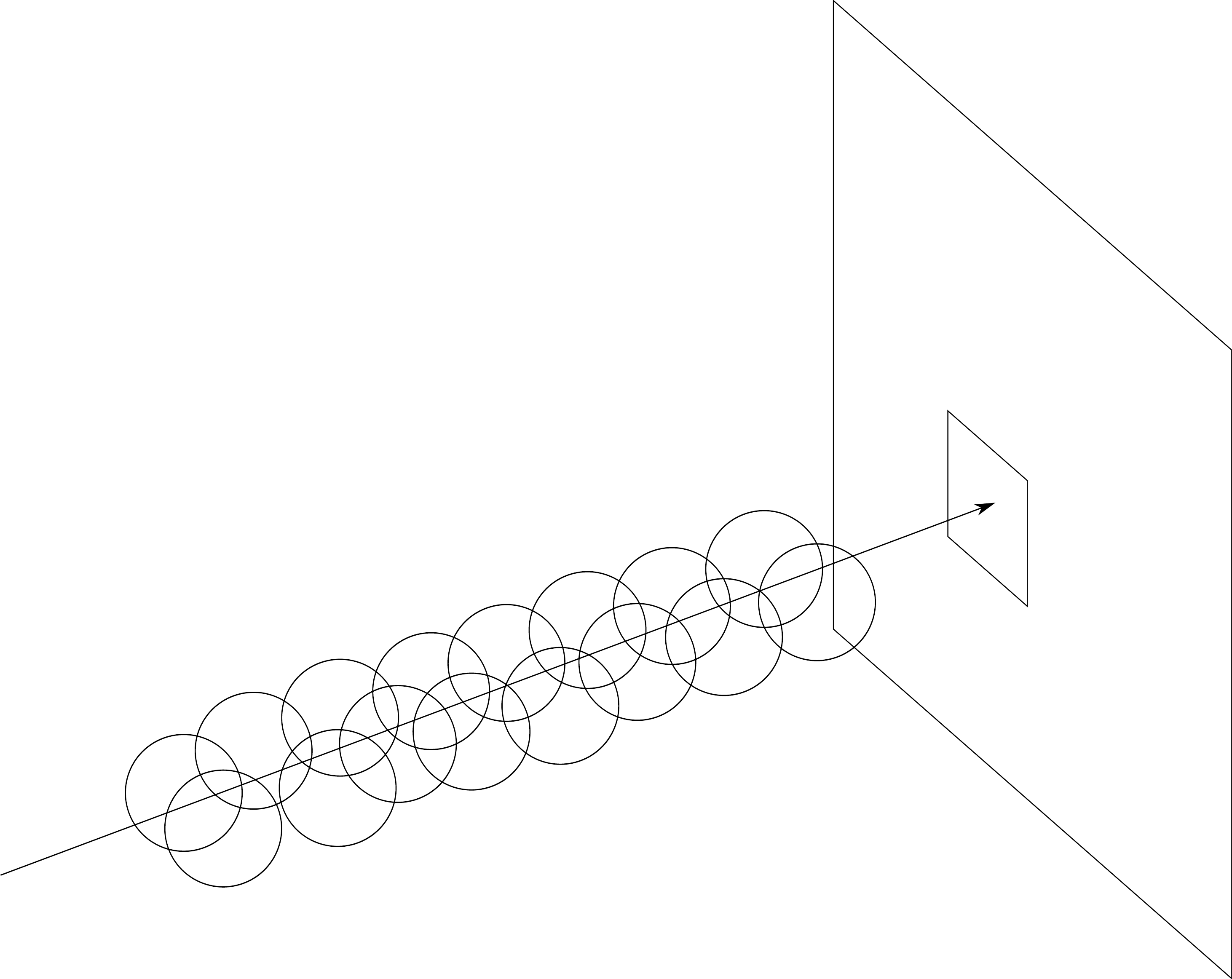}
  \caption{Schematic of a single pixel (small square) of an intensity map (large square). The ray passes through the centre of the pixel at an angle normal to the surface of the map.}
  \label{fig:intensity}
\end{figure}

We generate intensity maps of the dust emission via ray tracing. We construct a virtual rectangular screen facing the ensemble of particles. This screen is divided into $N_\textsc{pix}$ pixels. We construct $N_\textsc{pix}$ rays, each of which has infinite length and passes through the centre of a pixel with direction normal to the screen. A schematic of a single pixel is show in Fig. \ref{fig:intensity}\,. \changes{All intensity maps presented here have $N_\textsc{pix}=200\times200$\,.}

Each ray intersects $N_\textsc{ray}$ particles. These are sorted into descending order of distance from the pixel. The intensity of the pixel is given by $I_{N_\textsc{ray}}$. This is calculated by iteration:
\begin{equation}
  I_i=I_{i-1}\,\exp(-\chi_{\lambda i}\,\varsigma_i)+\frac{B_\lambda(T_i)\,\kappa_{\lambda i}}{\chi_{\lambda i}}[1-\exp(-\chi_{\lambda i}\,\varsigma_i)]\,,
  \label{eqn:ray_trace}
\end{equation}
where $i=1,2,\ldots,N_\textsc{part}$. In the case where $i=1$, $I_0$ is the background intensity.

The intensity from point sources (e.g. stars) may also be added to the map. This is achieved by (i) locating the pixel in which the source lies, (ii) calculating the optical depth $\tau$ between the source and the pixel and (iii) adding $L_\lambda\,\mathrm{e}^{-\tau}/(4\uppi\,A_\textsc{pix})$ to the pixel intensity. Here $L_\lambda$ is the monochromatic luminosity of the source and $A_\textsc{pix}$ is the area of the pixel.

\section{Example calculations}
\label{sec:examples}

\begin{figure}
\includegraphics[width=\columnwidth]{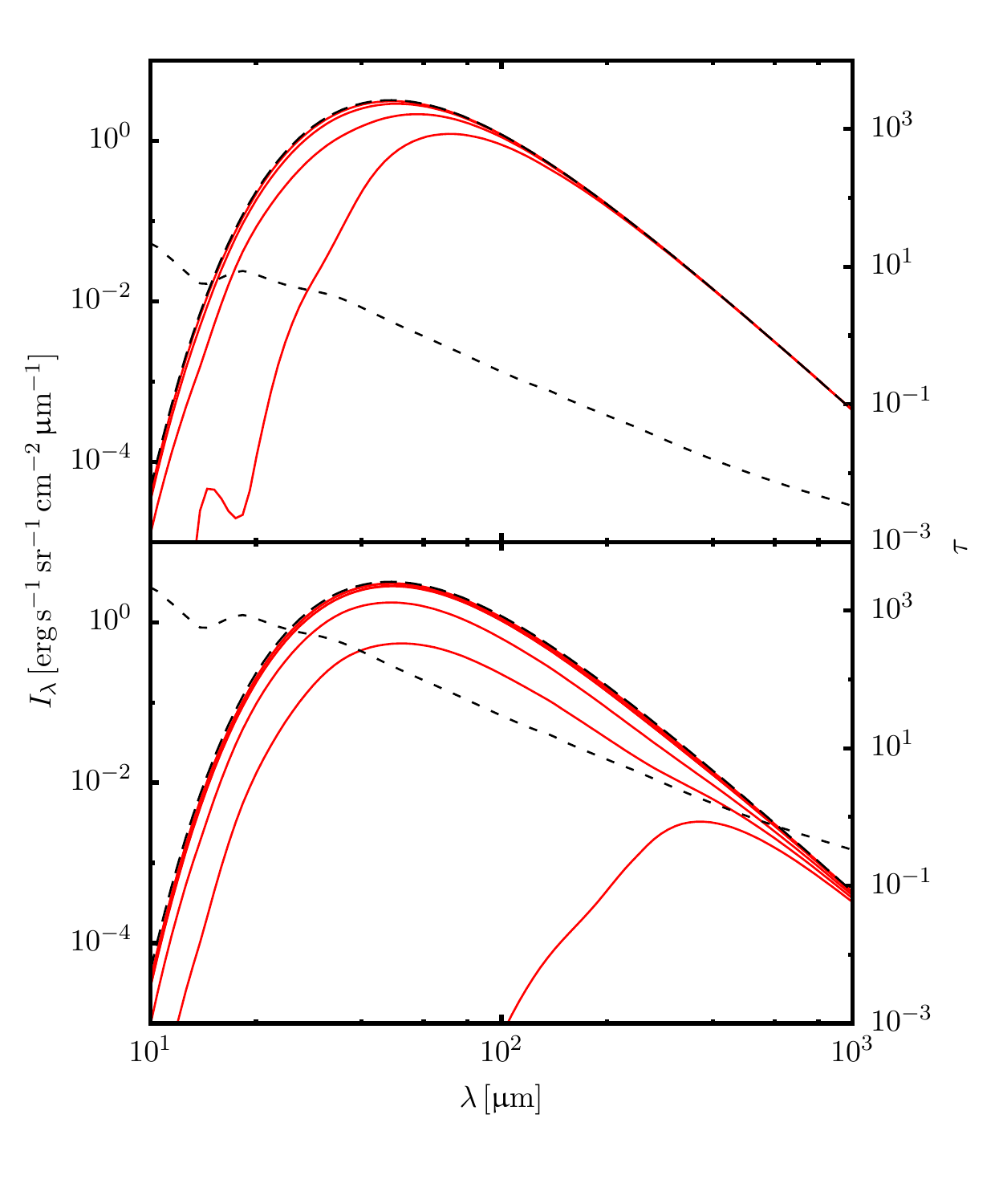}
\caption{The spectral energy distribution (SED) through the centre of the Bonnor-Ebert spheres, with $M=1\,\mathrm{M_\odot}$ (top) and $M=100\,\mathrm{M_\odot}$ (bottom). In both cases, the sphere is illuminated by an undiluted $60\,\mathrm{K}$ blackbody. The left $y$-axis gives the intensity calculated by successive iterations of the algorithm. The lowest red line shows the initial intensity, where each particle has a temperature of $10\,\mathrm{K}$. The lines above show the SED converging towards a $60\,\mathrm{K}$ blackbody (black long-dashed line). The right $y$-axis give the optical depth through the centre of the sphere (short-dashed black line).}
\label{fig:spec_conv}
\end{figure}

\begin{figure}
\includegraphics[width=\columnwidth]{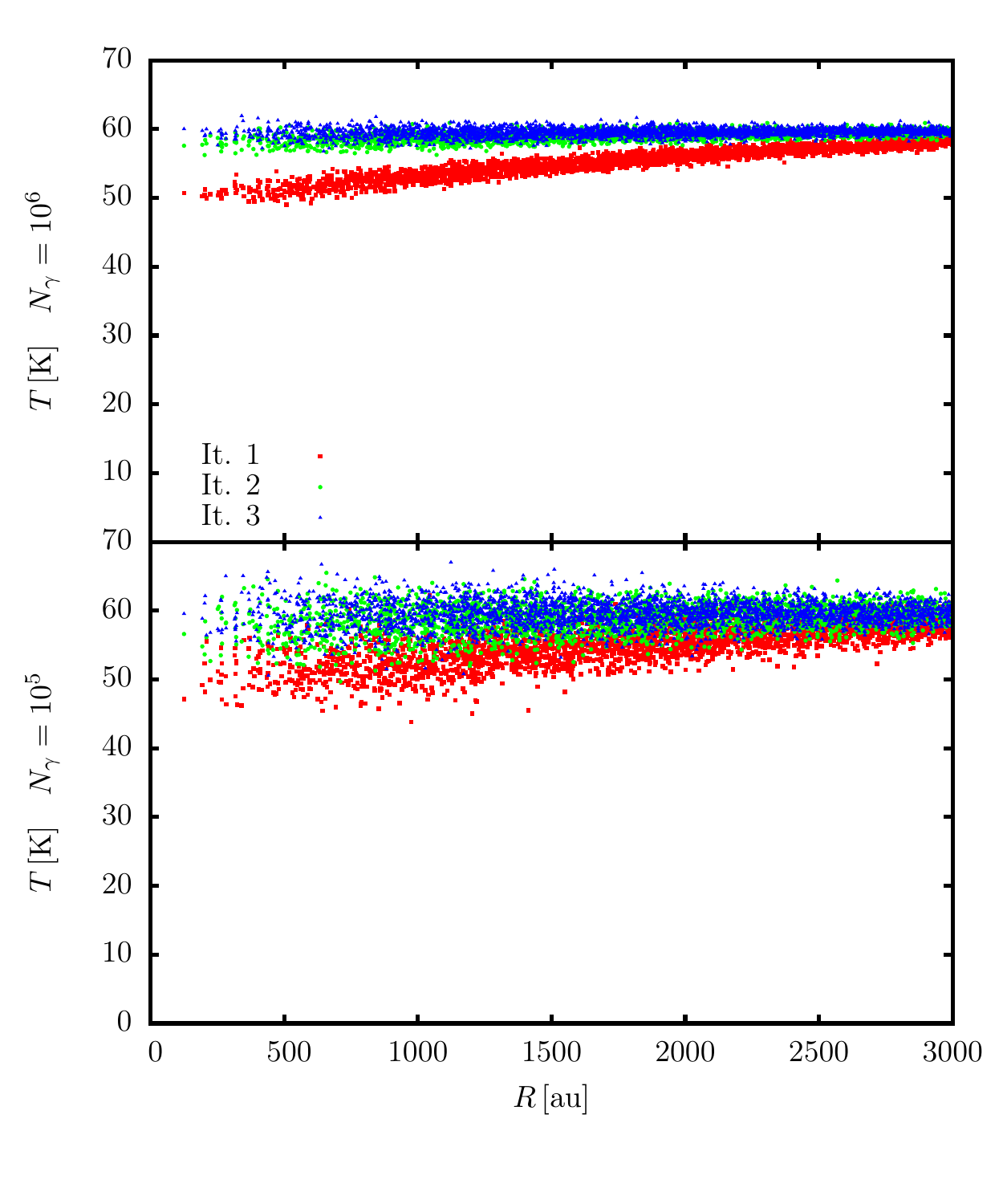}
\caption{Particle temperature versus radius for optically thin sphere. The top frame shows the temperature calculation with $10^6$ luminosity packets. The bottom frame shows the calculation repeated with $10^5$ luminosity packets. The particle temperatures from the first iteration are plotted as red squares, from the second iteration as green circles and from the third iteration as blue triangles.}
\label{fig:temp_conv}
\end{figure}

In the following calculations, we use the dust optical properties derived by \citep{LD01}. Here, the dust is a mixture of carbonaceous and amorphous silicate grains. The size distribution follows \citet{WD01} with $R_v=5.5$ and is normalised to give a dust-to-gas mass ratio of one percent. The code does not currently support imaging of scattered light (see \S \ref{sec:future}), so we set the albedo to zero, i.e. $\kappa_\lambda\equiv\chi_\lambda$.

We present two examples. The first is a benchmark which tests the ability of the algorithm to reach thermal equilibrium with a radiation field. In the second example, we use the outputs of an SPH simulation to generate spectra and intensity maps of an embedded multiple system.

\subsection{Blackbody radiation field}

\subsubsection{Undiluted blackbody radiation field}
\label{sec:sphere}

In an undiluted blackbody radiation field, the intensity $I_\lambda(\boldsymbol{x},\boldsymbol{n})$ is equal to the Planck function $B_\lambda(T_\textsc{bb})$. By construction, an absorbing/emitting object in this field has a surface intensity equal to $B_\lambda(T_\textsc{bb})$. Following Kirchhoff's law of radiation---a good absorber is an equally good emitter---the surface temperature must therefore be equal to $T_\textsc{bb}$. Furthermore, because the emission spectrum is identical to the absorption spectrum, the object is invisible with respect to the background intensity.

\subsubsection{Set up}

We set up a sphere of $N_\textsc{part}=3\times10^5$ particles. The sphere has radius $R=3000\,\mathrm{au}$ and the density profile of a critical Bonnor-Ebert sphere, truncated at dimensionless radius $\xi=6.451$\,. In the first instance, we give the sphere mass $M=1\,\mathrm{M_\odot}$ (optically thin), in the second $M=100\,\mathrm{M_\odot}$ (optically thick). We place the sphere in an undiluted $T_\textsc{bb}=60K$ blackbody radiation field. This is simulated by constructing a virtual shell which encompasses all the particles and their smoothing kernels, i.e. $R_\textsc{field}\gtrsim2000\,\mathrm{au}$. The shell is given luminosity $L_\textsc{field}=4\,\uppi\,R_\textsc{field}^2\,\sigma_\textsc{sb}\,T^4$. This luminosity is divided into $N_\gamma=10^6$ luminosity packets, which are directed inwards towards the particles. The wavelength of each packet is drawn randomly from the Planck function $B_\lambda(T_\textsc{bb})$. Each particle is given an initial temperature $T=10\,\mathrm{K}$ and \textsc{spamcart} is iterated until the mean change in temperature is less than $1\,\mathrm{K}$.

\subsubsection{Results}

Fig. \ref{fig:spec_conv} shows the spectral energy distribution (SED) though the centre of the sphere. In the optically thin case, we see that the SED converges on a $60\,\mathrm{K}$ blackbody after about three iterations. In the optically thick case, the SED converges after about five iterations.

\changes{
Fig. \ref{fig:temp_conv} shows the particle temperatures after the first three iterations of the algorithm with the optically thin sphere. This includes calculations with both $N_\gamma=10^6$ and $N_\gamma=10^5$. When $N_\gamma=10^6$, the particle temperatures settle on a narrow distribution of $T=59.7\pm0.4\mathrm{K}$. When $N_\gamma=10^5$ the temperatures distribution is $T=59.5\pm1.1\,\mathrm{K}$. This demonstrates the Poisson-like uncertainties on Monte Carlo calculations, i.e. a factor of $N$ fewer luminosity packets increases the signal-to-noise ratio by a factor of $\sqrt{N}$.
}

\subsubsection{Diluted blackbody radiation field}

We repeat the optically thick simulation, this time with a $120\,\mathrm{K}$ blackbody, diluted by a factor of $1/16$. In this instance, we expect the dust to reach local radiative equilibrium at some temperature $T<120\,\mathrm{K}$. Here, when a luminosity packet is absorbed by the dust, it is usually re-emitted at a longer wavelength. The sphere should therefore appear to glow, relative to the background, at long wavelengths and cast a silhouette at short wavelengths.

\begin{figure}
\hspace{0.5cm}\includegraphics[width=\columnwidth]{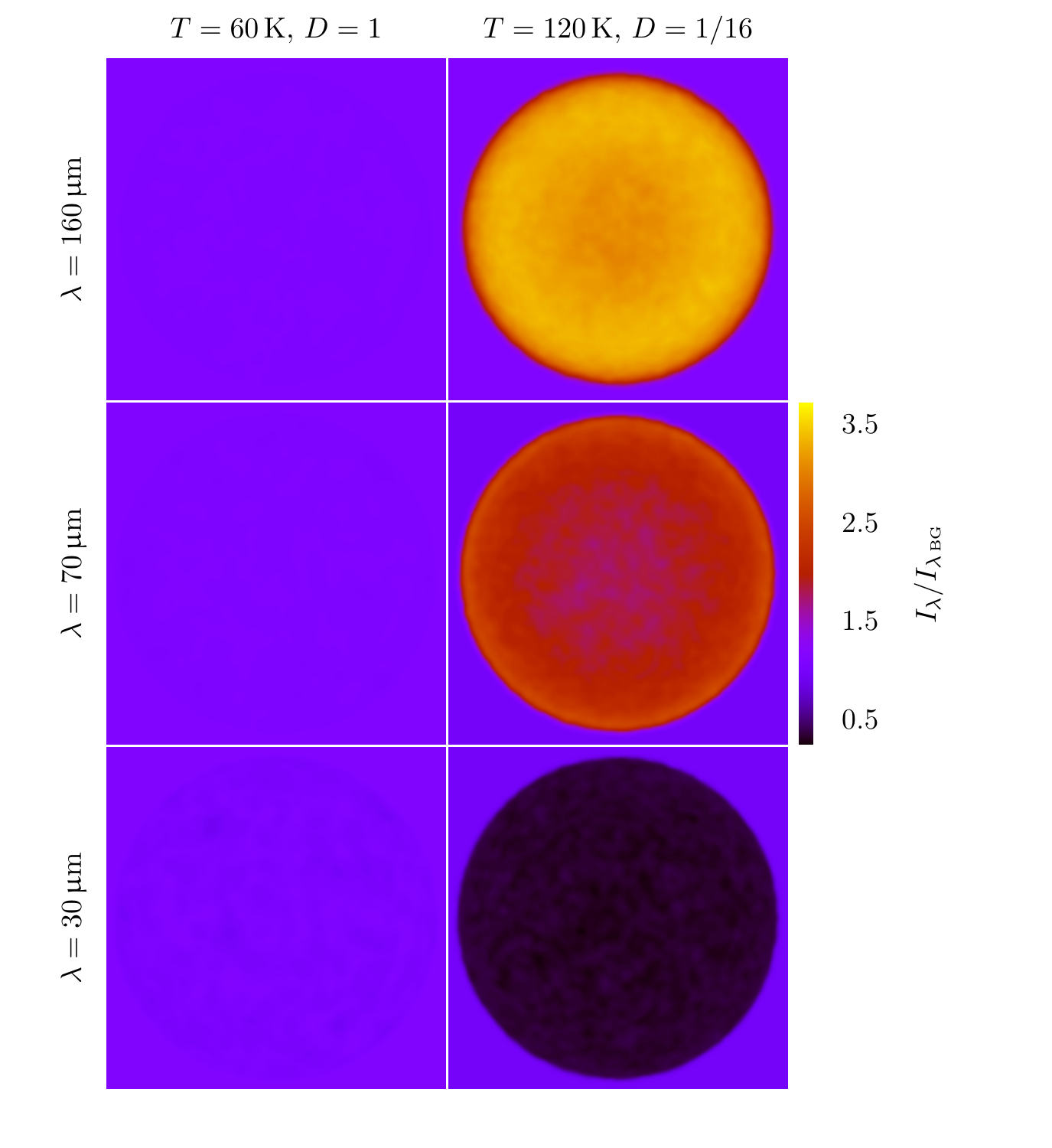}
\caption{Intensity maps of a sphere of $N_\textsc{part}=3\times10^5$ particles with $M=100\,\mathrm{M_\odot}$ and $R=2000\,\mathrm{au}$. The radial density profile follows that of a critical Bonnor-Ebert sphere. The left column shows the intensity when the sphere is illuminated by a $60\,\mathrm{K}$ blackbody radiation field. The right column shows the intensity when the sphere is illuminated by a $120\,\mathrm{K}$ blackbody radiation field, diluted by a factor of $1/16$. The top row gives the intensity at $\lambda=160\,\mathrm{\upmu m}$, the middle row gives the intensity at $\lambda=70\,\mathrm{\upmu m}$ and the top row gives the intensity at $\lambda=30\,\mathrm{\upmu m}$. The colour scale gives the intensity, normalised by the background intensity.}
\label{fig:equilibrium}
\end{figure}

Fig. \ref{fig:equilibrium} shows intensity maps of the optically thick sphere. The left column shows the results of the undiluted blackbody field, the right column shows the results for the diluted field. The sphere in the undiluted field is almost invisible relative to the background intensity at wavelengths between $30$ and $160\,\mathrm{\upmu m}$. There is some visible noise at $30\,\mathrm{\upmu m}$, of order ten percent. Here, the optical depth through the sphere is very high ($\tau\sim100$) and the intensity is very sensitive to temperature fluctuations across the the sphere's outer surface. The sphere in the diluted field behaves as expected; it is brighter than the background at $160\,\mathrm{\upmu m}$ and darker than the background at $30\,\mathrm{\upmu m}$.

\subsection{Embedded sextuple system}
\label{sec:sextuple}

\begin{figure*}
\centering
\includegraphics[width=0.8\textwidth]{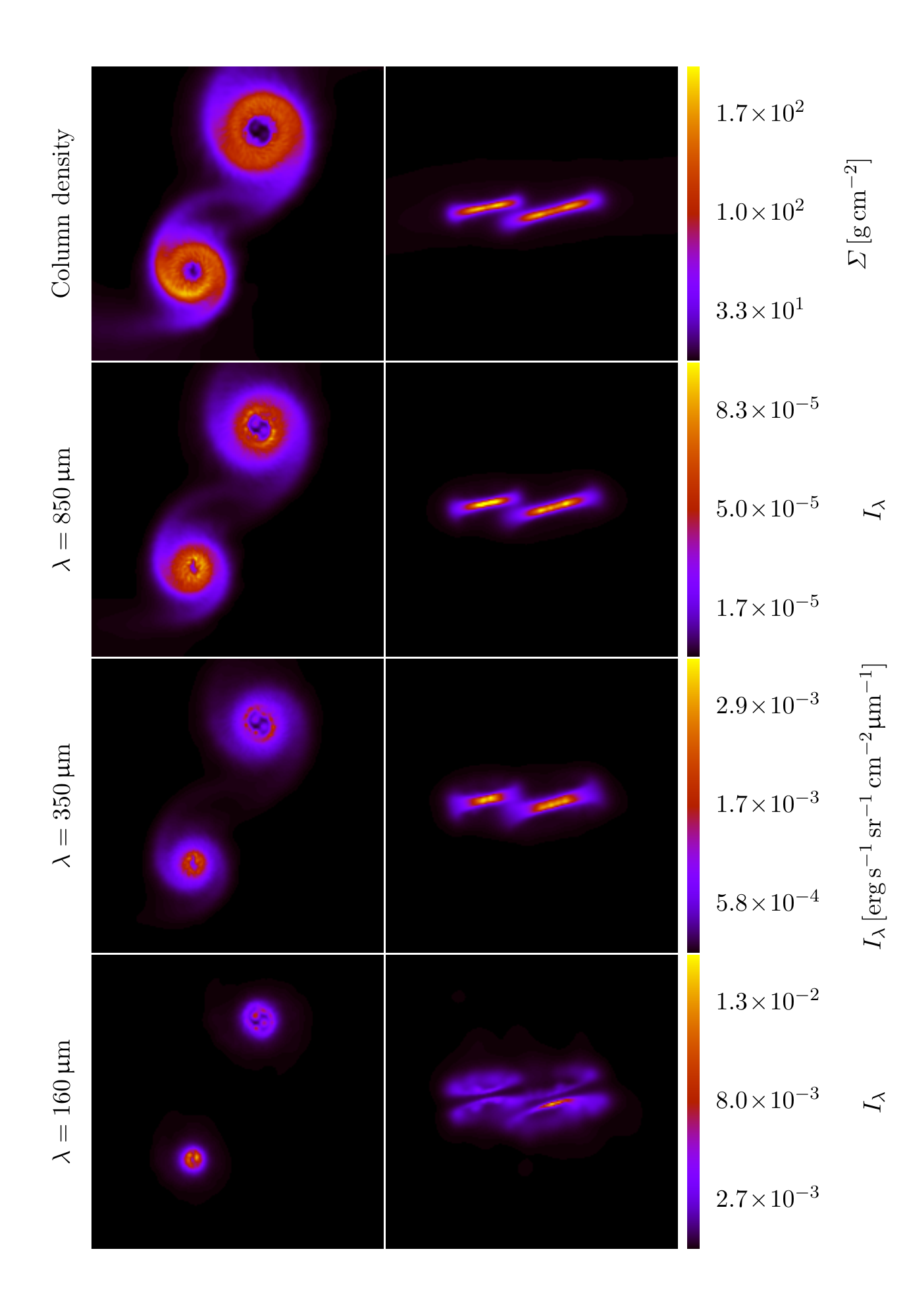}
\caption{Intensity maps of dust emission from a simulated protostellar sextuple system. The left column shows a face-on view, the right column shows shows an edge-on view. The top row shows the column density of the system. The following three rows show the intensity at $\lambda=850\,\mathrm{\upmu m}$, $350\,\mathrm{\upmu m}$  and $160\,\mathrm{\upmu m}$. Each frame has area $670\,\mathrm{au}\times670\,\mathrm{au}$. Note: the colour bar applies to the right column. Figures in the left column have been linearly scaled to fit in the same range.}
\label{fig:multi_img}
\end{figure*}

\begin{figure}
\includegraphics[width=\columnwidth]{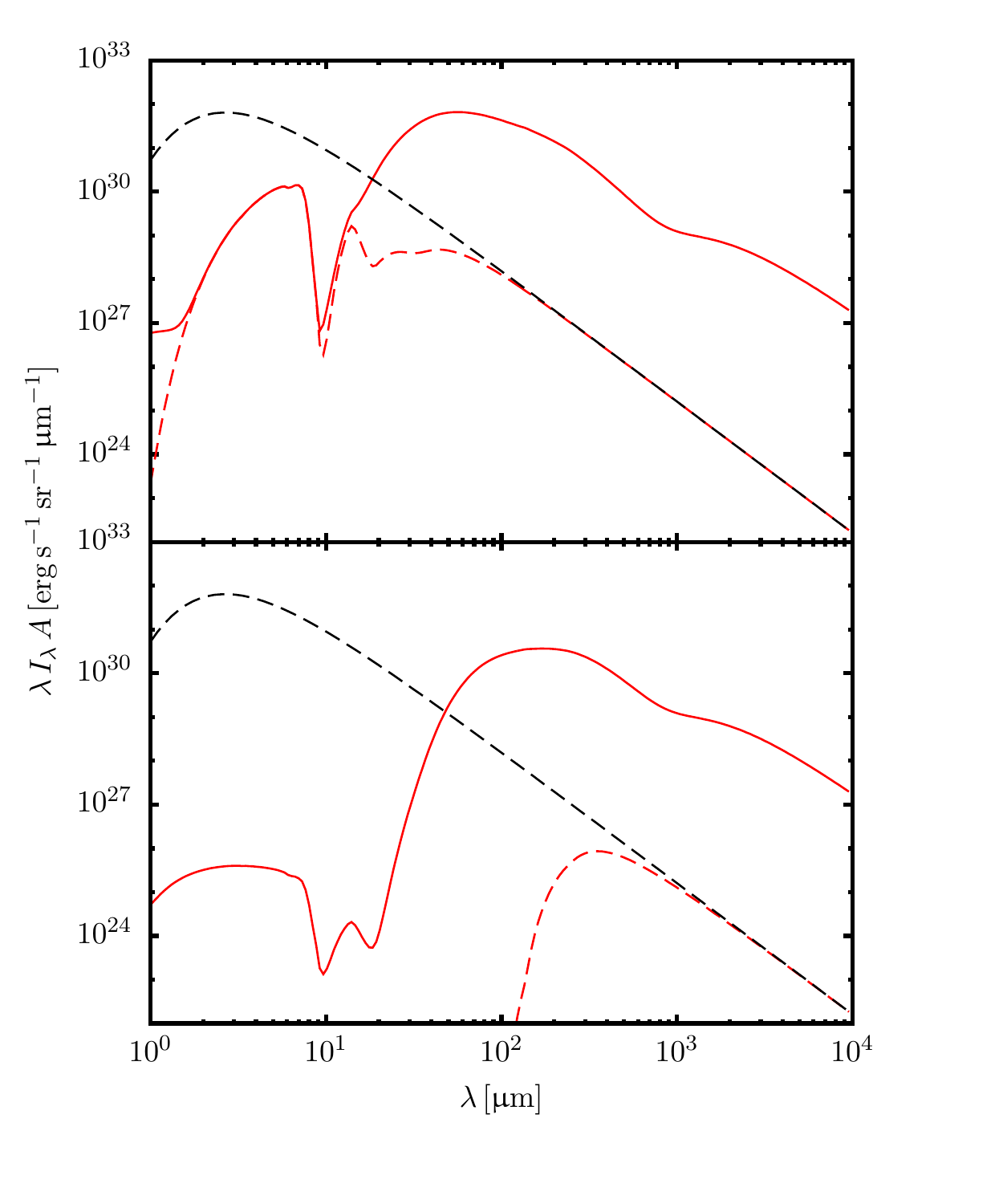}
\caption{Spectra of dust emission and starlight from the sextuple system. Values are integrated over the map area shown in Fig. \ref{fig:multi_img}\,. The top frame shows the spectrum from a face-on view. The bottom frame shows the spectrum from an edge-on view. The solid red lines show the total emission. The dashed red line shows the emergent emission from the six protostars. The dashed black line shows the protostellar emission in the absence of dust extinction.}
\label{fig:multi_spec}
\end{figure}

We demonstrate a more realistic application of the algorithm by calculating the dust emission from a protostellar multiple system. High order multiple systems, i.e. $N_\star\geq3$, are observed amongst mature field stars \citep[e.g.][]{T08,ET08}. These systems are more common amongst pre-Main Sequence stars, embedded in star forming regions such as Taurus and Ophiuchus \citep[e.g.][]{LZW93,RKL05,KIM11}. Furthermore, similar systems form routinely in simulations of molecular clouds and cores \citep[e.g.][]{DCB04b,B09a,LWHSW14,LWHSW14b}.

\subsubsection{Set up}

We take a snapshot from one of the SPH simulations presented by \citet{LWHSW14}\footnote{These simulations use initial conditions drawn from distribution functions that reproduce the observed properties of the cores in Ophiuchus. The simulation used here is No. 52, with episodic radiative feedback.}. Here, a $1.3\,\mathrm{M_\odot}$ core collapses and fragments into seven protostars (represented by sink particles). One of these is ejected from the core and the remaining six form a stable sextuple system. At $t=10^5\,\mathrm{years}$ (after the initial collapse), the protostars in the sextuple have masses between $0.08\,\mathrm{M_\odot}<M_\star<0.15\,\mathrm{M_\odot}$. The ejected protostar has mass $M=0.01\,\mathrm{M_\odot}$. The remainder of the initial mass ($M_\textsc{gas}=0.65\,\mathrm{M_\odot}$, $N_\textsc{part}=6.5\times10^4$) is distributed between discs and and a diffuse envelope.

We irradiate the system with an undiluted $2.73\,\mathrm{K}$ blackbody radiation field and a $10000\,\mathrm{K}$ blackbody field, diluted by a factor of $5\times10^{-15}$. This represents contributions from the Cosmic Microwave Background (CMB) and the galactic stellar population. The protostars are treated as blackbody sources with $R=4\,\mathrm{R_\odot}$ and $1100\,\mathrm{K}<T<1500\,\mathrm{K}$; these temperatures are estimated using the \citet{SWH11} episodic accretion model.

\subsubsection{Intensity maps and spectra}

Fig. \ref{fig:multi_img} shows column density and intensity maps of the embedded system. The system is viewed both face-on and edge-on. In the face-on view, we see a balanced quadruple system (top right) in orbit with a binary system (bottom left), separated by $S\sim400\,\mathrm{au}$.

In both face-on and edge-on views, the intensity at $850$ and $350\,\mathrm{\upmu m}$ roughly traces the column density. At $160\,\mathrm{\upmu m}$, face-on, only the the inner regions of the discs are visible. When viewed edge-on, the discs are opaque and significant intensity is only seen above and below their midplanes.

Fig. \ref{fig:multi_spec} shows the emission spectrum of the system. When viewed face on, we may define different wavebands, dominated by different sources of radiation. Between $10^3\,\mathrm{\upmu m}<\lambda\leq10^4\mathrm{\upmu m}$, the dust is optically thin and most of the radiation is from the CMB. Between $10\,\mathrm{\upmu m}<\lambda\leq10^3\mathrm{\upmu m}$, most of the radiation is from dust emission. Starlight is largely unattenuated in these wavebands, but its fraction of the total emission is very low. At wavelengths less than $10\,\mathrm{\mu m}$, the majority of the emission is attenuated starlight. Similar dust emission is seen when the system is viewed edge on. However, the starlight is almost completely extinguished at $\lambda<200\,\mathrm{\mu m}$. We note that the level of extinction at $\lambda<10\,\mathrm{\mu m}$ is probably exaggerated as we have not captured scattered light in these spectra. Nevertheless, in the face-on view, we still see the double peaked spectra typical of protostellar discs \citep[e.g.][]{R11,WRB13}.

\section{Future developments}
\label{sec:future}

\subsection{Performance}

\changes{
While the code is not yet heavily optimized, the algorithm is efficient enough to run on an Intel i5-4200U dual core CPU at $1.60\,\mathrm{GHz}$ in a reasonable time. A single iteration of the optically thick sphere in \S\ref{sec:sphere} requires $3.5\times10^{-4}\,\mathrm{s}$ per packet per CPU core. The optically thin sphere requires $5.0\times10^{-5}\,\mathrm{s}$ per packet per CPU core. A single iteration of the sextuple system in \S\ref{sec:sextuple} requires approximately $1.5\times10^{-5}\,\mathrm{s}$ per packet per CPU core. We note that the optimal number of packets $N_\gamma$ is not obvious. In this paper, we select sensible values through a process of trial and error.
}

\changes{
A thorough analysis of the processing time required for this algorithm in different scenarios is beyond the scope of this paper. However, we can estimate the time scaling for simple systems such as a star embedded in a uniform density sphere. Here,
\begin{equation}
  t( \bar{\tau},N_\gamma,\xi,N_\textsc{sph})\propto \bar{\tau}^2\,N_\gamma\,\xi^2\,N_\textsc{sph}^{1/3}\,\log N_\textsc{sph}.
\end{equation}
The flight path of a packet is proportional to the square of the average optical depth through the system $\bar{\tau}$. The number of particles along the path is proportional to the square of the kernel extent $\xi$ and the cube root of the total number of particles $N_\textsc{sph}$. The time taken to find each particles scales roughly with the log $N_\textsc{sph}$.
}

\changes{
One may ensure that each particle is visited the roughly same number of times, regardless of $N_\textsc{sph}$, by setting $N_\gamma\propto N_\textsc{sph}^{2/3}$. Ignoring $\xi$, the time scaling is now $t(\tau,N_\textsc{sph})\propto\bar{\tau}^2\,N_\textsc{sph}\,\log N_\textsc{sph}$. This is similar to that of an SPH simulation timestep. Therefore it may be feasible in the future to run this algorithm on-the-fly in an SPH simulation, especially in circumstances where $\bar{\tau}\lesssim 1$\,.
}

\subsection{Features}

At present, we are able to generate intensity maps of the dust emission, plus attenuated light from background and point sources. However, it is desirable to also capture light scattered by dust. This can be achieved by modifying Eqn. \ref{eqn:ray_trace}. Here, the term $B_\lambda(T)\kappa_\lambda$ represents the amount of dust emission per unit solid angle, per unit wavelength, per unit mass. We may add an extra term which accounts for scattered light:
\begin{equation}
  B_\lambda(T)\kappa_{\lambda}\to B_\lambda(T)\kappa_{\lambda}+\frac{1}{\Delta\lambda}\frac{\varepsilon_0}{\Delta t}\frac{1}{m}\sum\limits_{j,\lambda,\Delta\lambda}\varPhi(\theta_j)\sigma_{\lambda j}\,\varsigma_j\,.
\end{equation}
Here, $\sigma_\lambda\equiv a_\lambda\,\chi_\lambda$ is the mass scattering coefficient and $\theta_j$ is the angle between luminosity packet $j$ and the viewing angle. The sum is only over luminosity packets with wavelengths in the interval $(\lambda,\lambda+\Delta\lambda)$. The scattering term must be calculated during the Monte Carlo iteration, so the viewing angle and wavelength interval must be known beforehand.

Other future developments include use of the partial diffusion approximation and modified random walk for regions of high optical depth \citep{MDD09}. The \textsc{spamcart} code will shortly be made open-source and uploaded to the \href{https://github.com/}{GitHub} repository.

\section{Summary}
\label{sec:summary}

We have developed a method for performing MCRT calculations directly on a distribution of SPH particles The algorithm operates differently from uniform-density cell methods, but the two schemes are mathematically equivalent. This allows an MCRT calculation to be performed on an SPH snapshot with (i) no loss in density resolution and (ii) no introduction of noise from mapping particles to cells.

We present a version of this algorithm that uses the \citet{Lucy99} method to compute (i) the propagation of luminosity packets through a medium and (ii) the radiative equilibrium temperature. The trajectories of the packets rely on the temperature of the medium so the calculation must be solved by iteration.

We provide two example calculations using the smoothed particle MCRT method. First, we show that a cloud is invisible when it is bathed in an undiluted blackbody radiation field. This holds for both  optically thin and optically thick cases. Therefore the code obeys Kirchhoff's law of thermal radiation. We note that this is a powerful test of any radiative transfer code and can be applied to any configuration. Conversely, a cloud which has reached radiative equilibrium with a diluted blackbody field glows at long wavelengths and casts a silhouette at short wavelengths. Second, we generate intensity maps and spectra of protostellar and dust emission from an embedded sextuple system. Here, a double peaked disc spectrum is seen when the system is observed face on. When viewed edge on, the opacity of the disc blocks nearly all of the starlight.

Future additions to the code include the addition of scattered light to synthetic observations and optimisations for optically thick regions of dust. The code, written in Fortran 2003/08 with OpenMP parallelisation, will be made publicly available in the near future.

\section*{Acknowledgements}

OL and APW gratefully acknowledge the support of a consolidated grant (ST/K00926/1) from the UK STFC. We also thank the referee for their considerate and constructive comments.

\bibliographystyle{mn2e}
\bibliography{refs}

\appendix

\section{Gridding Errors}
\label{apn:gridding}

\begin{figure}
  \includegraphics[width=\columnwidth]{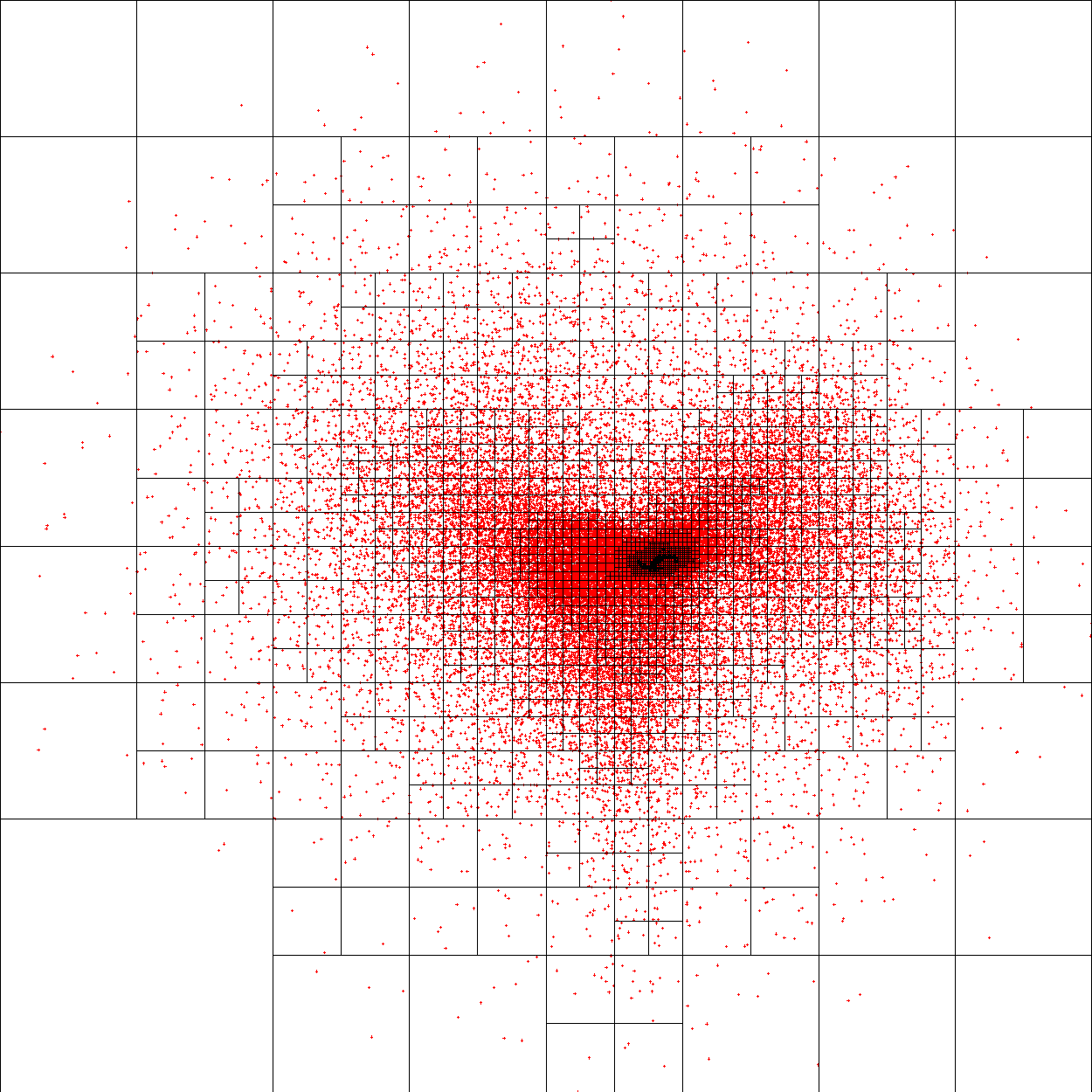}
  \caption{An $x$-$y$ projection of the octree and particle ensemble with $N_\textsc{leaf}=8$. The root cell is roughly cubic with an edge length of $0.2\,\mathrm{pc}$. The smallest leaf cells have an edge-length of $2.4\times10^{-5}\,\mathrm{pc}$.}
  \label{fig:octree}
\end{figure}

Transposing an ensemble of SPH particles onto a grid introduces errors to the density field. Here, we perform a brief analysis on these errors. We generate an octree around the distribution of particles used to model the sextuple system in \S\ref{sec:sextuple}. First, we build the smallest cuboidal \textit{root} cell that contains all of the particles. This cell is recursively subdivided into eight equal volume cells until each \textit{leaf} cell contains $N_\textsc{leaf}$ particles or fewer. Fig. \ref{fig:octree} shows the $x$-$y$ projection of the particles and the octree with $N_\textsc{leaf}=8$\,.

\begin{figure}
  \includegraphics[width=\columnwidth]{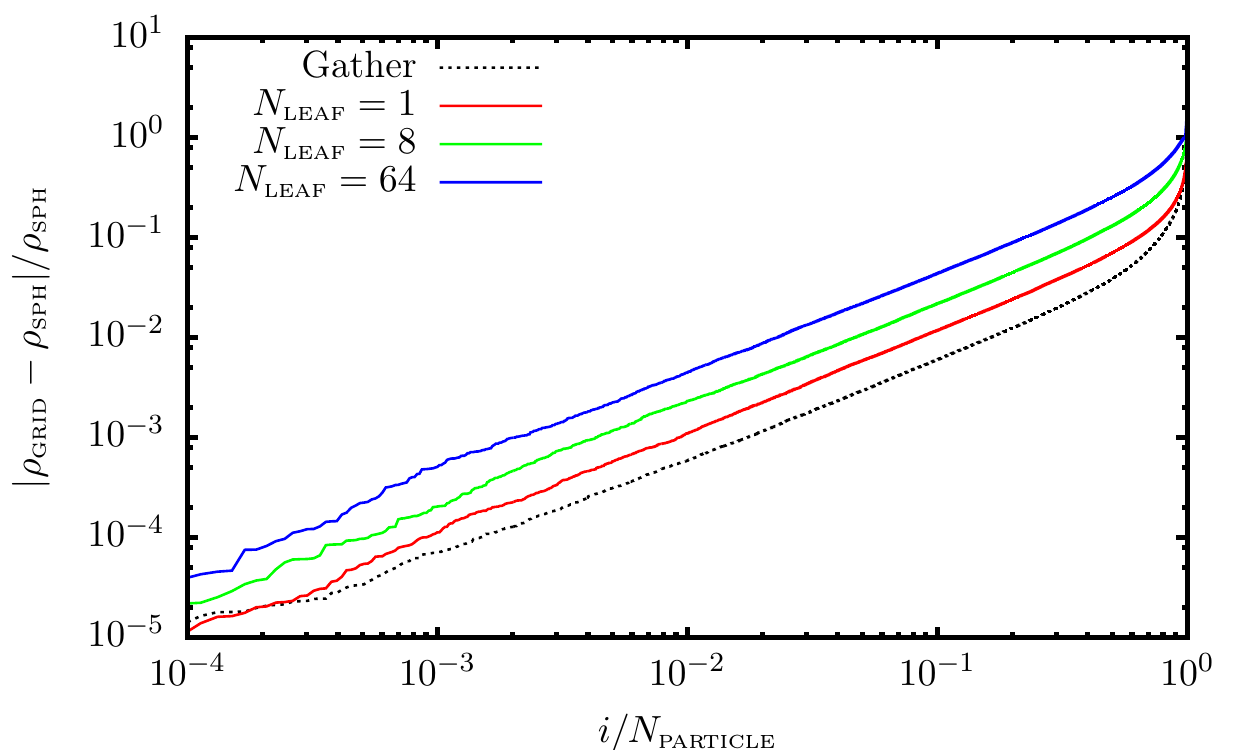}
  \caption{The relative difference between the grid cell density and the SPH particle density against particle number. Particles are sorted in order of ascending error. From top to bottom, the solid coloured lines show $N_\textsc{leaf}=64$, $8$ and $1$. The dotted black line shows the relative difference between the gather and scatter density for each particle, i.e. $|\rho_\textsc{gather}-\rho_\textsc{scatter}|/\rho_\textsc{scatter}$.}
  \label{fig:density_error}
\end{figure}

The density of each cell $\rho_\textsc{grid}$ is computed using an SPH \emph{scatter} calculation (see Eqn. \ref{eqn:scatter}) at the centre of the cell. We compare this with the particle density $\rho_\textsc{sph}$ (scatter calculation) of each particle in the cell. Fig. \ref{fig:density_error} shows the relative difference between $\rho_\textsc{sph}$ and $\rho_\textsc{grid}$ for the ensemble of particles with different values of $N_\textsc{leaf}$. Note that the particles are sorted into ascending order of density error. For comparison, we also include the relative difference between the particle scatter density and its \textit{gather} density (see Eqn. \ref{eqn:gather})\footnotemark. In Table \ref{tab:density_error} we give some further statistics including the number of leaf cells, the maximum depth of the tree, the total mass of the grid (defined as the sum of the products of leaf cell volume and leaf cell density) and the 90th centile density error.

\footnotetext{The difference between the scatter and gather density calculations does not indicate that SPH is somehow wrong. It reflects the difference between the mass to volume ratio of an SPH particle (gather), and the superposition of multiple smoothing kernels at a given point in space (scatter).}

Gridding the density field with $N_\textsc{leaf}=$ 1, 8 and 64 incurs a 90th centile density error of 24\%, 45\% and 76\% respectively. Furthermore, the mass of the grid is roughly 80\% to 90\% that of the particle ensemble. We note that although the method presented in the paper is more accurate than gridding an ensemble of particles, MCRT calculations on octrees are mathematically and computationally simpler.

\begin{table}
  \begin{tabular}{cccccc}
    \hline
    $N_\textsc{leaf}$ & $N_\textsc{oct}$ & $N_\textsc{fixed}$ & Levels & $M_\textsc{grid}/M_\textsc{sph}$ & $\mathrm{Error}_{90\%}$ \\
    \hline
    1 & 154281 & $65536^3$ & 16 & 0.80 & 0.24 \\
    8 & 22632 & $8192^3$ & 13 & 0.87 & 0.45 \\
    64 & 3242 & $4096^3$ & 12 & 0.88 & 0.76 \\
    Gather & -- & -- & -- & -- & 0.18 \\
    \hline
  \end{tabular}
  \caption{Statistics of the particle octrees. The first column give the maximum number of particles per leaf. The second column gives the total number of leaf cells. The third column gives the number of cells in an equivalent regular grid. The forth column gives the ratio of the total grid mass to the total particle mass. The fifth column gives the 90th centile density error (see Fig. \ref{fig:density_error}).}
  \label{tab:density_error}
\end{table}

\section{Kernel function}
\label{apn:kernel}

The M4 kernel has support $s\leq2$, where $s=r/h$. The density at $s$ is given by
\begin{equation}
  w(s)=\frac{1}{\uppi}\left(\frac{1}{4}\max(2-s,0)^3-\max(1-s,0)^3\right)\,,
\end{equation}
and the density gradient is given by
\begin{equation}
  \frac{\mathrm{d}}{\mathrm{d}s}w(s)=-\frac{1}{\uppi}\left(\frac{3}{4}\max(2-s,0)^2-3\max(1-s,0)^2\right)\,.
\end{equation}
The column density from dimensionless impact parameter $c=b/h$ to $s$ is given by
\begin{equation}
  W(c,s)=\frac{1}{\uppi}
    \begin{cases}
      W_\textsc{inner}(c,s), & c\leq1,\,s\leq1,\\
      W_\textsc{inner}(c,1)+W_\textsc{outer}(c,s), & c\leq1,\,s>1,\\
      W_\textsc{outer}(c,s), & c>1,\,s>1\,,\\
      0, & c>2\,,
    \end{cases}
  \label{eqn:w_col}
\end{equation}
where $c\leq s$ and
\begin{equation}
  \begin{split}
    W_\textsc{inner}&(c,s)=\sqrt{s^2-c^2}\\
    &\times\left(\frac{3}{16}s^3-\frac{1}{2}s^2+\frac{9}{32}c^2s-c^2-1\right)\\
    &+\frac{9}{32}c^4\log\left(\frac{s+\sqrt{s^2-c^2}}{c}\right)\,,\\
    W_\textsc{outer}&(c,s)=\sqrt{s^2-c^2}\\
    &\times\left(-\frac{1}{16}s^3+\frac{1}{2}s^2-\frac{3}{32}(c^2+16)\,s+c^2+2\right)\\
    &-\frac{3}{32}(c^2+16)\,c^2\log\left(\frac{s+\sqrt{s^2-c^2}}{c}\right)\,.
  \end{split}
\end{equation}
In practice, Eqn. \ref{eqn:w_col} can either be computed on-the-fly or stored in a triangular lookup table in the the range \mbox{$0\leq c^2\leq4$} and \mbox{$c^2\leq s^2\leq4$}\,.

\section{k-d tree}
\label{apn:tree}

We use a $k$-d tree (in three dimensions) to find the particles intersected by a ray. We construct the tree by placing the entire ensemble of particles into a \emph{root} cell. The root cell contains two \emph{branch} cells. We select the dimension $k$ across which the particle position $\boldsymbol{x}$ has the greatest variance. All particles with $x_k$ up to and including the median value are placed in the first cell and the remaining particles are placed in the second cell. This process is repeated recursively until each cell contains 8 or fewer particles, in which case they are \emph{leaf} cells. For each cell, we calculate the Axis-Aligned Bounding Box (AABB) that encompasses all of the particle smoothing volumes within the cell.

We find the particles intersected by a ray by using the slab method \citep[e.g.][]{WBMS05}. Consider a ray with origin $\boldsymbol{o}$, direction $\boldsymbol{n}$ and length $l$. Also, consider an AABB with lower limits $\boldsymbol{b}^\textsc{min}$ and upper limits $\boldsymbol{b}^\textsc{max}$. Along each dimension $k$, we define a slab (a slab is a space between two parallel planes) with lower limit $b^\textsc{min}_k$ and upper limit $b^\textsc{max}_k$. The length of the ray segment, $t^\textsc{max}_k$-$t^\textsc{min}_k$, within the slab can be calculated:
\begin{equation}
  \begin{split}
    t^\textsc{min}_k&=\min([b^\textsc{min}_k - o_k]/n_k,\,[b^\textsc{max}_k - o_k]/n_k)\,,\\
    t^\textsc{max}_k&=\max([b^\textsc{min}_k - o_k]/n_k,\,[b^\textsc{max}_k - o_k]/n_k)\,\,.
  \end{split}-
\end{equation}
The length of the ray through the AABB in all three dimensions, $T_\textsc{max}-T_\textsc{min}$, is given by
\begin{equation}
  \begin{split}
  T_\textsc{min}&=\max({t^\textsc{min}_1,t^\textsc{min}_2,t^\textsc{min}_3})\,,\\
  T_\textsc{max}&=\min({t^\textsc{max}_1,t^\textsc{max}_2,t^\textsc{max}_3})\,.
  \end{split}
\end{equation}
The ray intersects the AABB if the following statements are true:
\begin{equation}
  \begin{split}
    T_\textsc{max}&>T_\textsc{min}\,,\\
    T_\textsc{min}&<l\,,\\
    T_\textsc{max}&>0\,.
  \end{split}
\end{equation}

Starting at the root cell, we check for an AABB-ray intersection. If the result is true, we open the branch cells and recursively repeat the process until we encounter a leaf cell. If the leaf cell is intersected by the ray, the cell's contents are added to a particle list. This method is not exclusive to $k$-d trees and may be used with any axis aligned tree, such as an octree \citep[e.g.][]{BH86}.

\section{Root-finding method}
\label{apn:roots}

Newton's method can be used to find the length of a luminosity packet trajectory. The equation $\varSigma(l)=\tau\,\chi_\lambda^{-1}$, where $\varSigma(l)$ is the column density, can be solved rapidly by iterating
\begin{equation}
  l_{n+1}=l_n-\frac{\varSigma(l_n)-\tau\,\chi_\lambda^{-1}}{\varSigma'(l_n)}\,,
  \label{eqn:lin_interp}
\end{equation}
where $\varSigma'(l_n)=\rho(\boldsymbol{o}+l_n\boldsymbol{n})$. This performs well if $l_n$ is near the root of $\varSigma(l)-\tau\,\chi_\lambda^{-1}$, but may fail to converge for poor initial choices of $l_n$.

We modify this method so that convergence is guaranteed. First we bracket the root, i.e. find values $l_{n-1}$ and $l_{n}$ so that $\varSigma(l_n)$ and $\varSigma(l_{n-1})$ are on opposite sides of $\tau\,\chi_\lambda^{-1}$. We assume that Eqn. \ref{eqn:length_zero} succeeds in overestimating $l$ and set $l_{n-1}=0$ and $l_n=l_0$. We now define the parabolic curve which (i) passes through point $[l_n,\varSigma(l_n)]$ with gradient $\varSigma'(l_n)$ and (ii) passes through point $[l_{n-1},\varSigma(l_{n-1})]$. This curve is described by the equation
\begin{equation}
  \begin{split}
  \varSigma(l_n)&-\varSigma(l_{n-1})\\
  &+\varSigma'(l_n)(l_{n-1}-l_n)\\
  &+\frac{1}{2}\varSigma''(l_n)(l_{n-1}-l_n)^2=0\,,
  \end{split}
\end{equation}
and by construction must have a single root in the interval $(l_{n-1},l_n)$. Rearranging for the second derivative,
\begin{equation}
  \varSigma''(l_n)=2\left[\frac{\varSigma(l_{n-1})-\varSigma(l_{n})}{(l_{n-1}-l_n)^2}-\frac{\varSigma'(l_n)}{l_{n-1}-l_n}\right]\,.
\end{equation}
Solving the quadratic formula, the iteration function is now
\begin{equation}
  l_{n+1}=l_n-\frac{2\,[\varSigma(l_n)-\tau\,\chi_\lambda^{-1}]}{\varSigma'(l_n)+\sqrt{[\varSigma'(l_n)]^2-2\,[\varSigma(l_n)-\tau\,\chi_\lambda^{-1}]\,\varSigma''(l_n)}}\,.
  \label{eqn:quad_interp}
\end{equation}
If $\varSigma(l_n)$ and $\varSigma(l_{n+1})$ are on opposite sides of $\tau\,\chi_\lambda^{-1}$, the new $l_{n-1}$ is set to the old $l_n$. Otherwise $l_{n-1}$ remains the same.

\label{lastpage}

\end{document}